\documentclass[draft]{agujournal}

\journalname{JGR-Atmospheres}

\begin{document}

\title{The impact of stochastic physics on climate sensitivity in EC-Earth}

\authors{K. Strommen\affil{1}, P. A. G. Watson\affil{1}, T. N. Palmer\affil{1}}

\affiliation{1}{Department of Physics, University of Oxford, Oxford, United Kingdom}

\correspondingauthor{K. Strommen}{Oxford University, Clarendon Laboratory, AOPP, Parks Road, OX1 3PU, Oxford. E-mail: kristian.strommen@physics.ox.ac.uk}



\begin{keypoints}
\item The inclusion of a stochastic scheme reduces climate sensitivity in a general circulation model.
\item This reduction, of around 10\%, is linked to changes in cloud cover and cloud optical depth feedbacks.
\item Well calibrated stochastic schemes may give more accurate global warming projections.
\end{keypoints}

%


\begin{abstract}
Stochastic schemes, designed to represent unresolved sub-grid scale variability, are frequently used in short and medium-range weather forecasts, where they are found to improve several aspects of the model. In recent years, the impact of stochastic physics has also been found to be beneficial for the model's long term climate. In this paper, we demonstrate for the first time that the inclusion of a stochastic physics scheme can notably affect a model's projection of global warming, as well as its historical climatological global temperature. Specifically, we find that when including the `stochastically perturbed parametrisation tendencies' scheme (SPPT) in the fully coupled climate model EC-Earth v3.1, the predicted level of global warming between 1850 and 2100 is reduced by 10\% under an RCP8.5 forcing scenario. We link this reduction in climate sensitivity to a change in the cloud feedbacks with SPPT. In particular, the scheme appears to reduce the positive low cloud cover feedback, and increase the negative cloud optical feedback. A key role is played by a robust, rapid increase in cloud liquid water with SPPT, which we speculate is due to the scheme's non-linear interaction with condensation.
\end{abstract}

%
%

\section{Introduction}
\label{sec:intro}

Estimating the extent of global warming due to anthropogenic forcing is one of the primary challenges in climate science, and, arguably, one of the most pressing problems to address for society as a whole. However, despite significant efforts in model development, estimates from state of the art climate models have remained relatively unchanged since the original IPCC report (\citep{Sherwood2014}). The majority of the spread in climate sensitivity across models can be attributed to differences in the response of clouds to an increase in greenhouse gases (\citep{Dufresne2008}), with low-level cloud response being particularly crucial (\citep{Zelinka2012a, Bony2005}). Because the equations of physics that modulate the hydrological cycle, and, therefore, cloud cover, are known to first order (\citep{Palmer2016}), variations across the models' representation of clouds are due to different choices in how to truncate these equations to a finite resolution. Of particular importance here are the simplified parameterisations used to determine the contribution from the sub-grid scale physics. Because the sub-grid scale contribution is not uniquely constrained by the gridscale state, there remains considerable amounts of choice involved in implementing these. While parameterisations have become increasingly sophisticated, they are still the dominant source of uncertainty in climate models.

One source of simulation error in sub-grid scale processes is the fact that a large portion of the variability at these scales is inherently unpredictable (\citep{Cohen2006, Davies2013}). In medium-range and seasonal forecasts using numerical weather prediction models, the use of stochastic schemes has become widespread as a means to sample this variability. Studies have shown that, when properly calibrated, such schemes have a beneficial impact on both the spread and mean state of these forecasts \citep{Weisheimer2011,Berner2017,Leutbecher2017}. In recent years, there has also been increasing interest in understanding the impact of these schemes on the long-term climate of a model. In \citep{Palmer2012}, it was argued that introducing stochasticity into climate models may be a key step towards eliminating persistent model biases and reducing uncertainty in climate projections, a view corroborated further by \citep{Christensen2019}. Since then, the insertion of a stochastic component into a climate model has been demonstrated to improve several key processes, including the El Ni\~no-Southern Oscillation \citep{Christensen2017a, Berner2018}, the Madden Julian Oscillation \citep{Wang2016} and the representation of the Indian monsoon \citep{Strommen2018}. Improvements were also found on regime behaviour, northern hemispheric blocking patterns and tropical precipitation \citep{Dawson2015,Davini2017,Watson2017}. Most of these studies focused on a particular, multiplicative noise scheme called the `stochastically perturbed parametrisation tendencies' scheme (SPPT: see Section \ref{sec:sppt_descr}). A more flexible variant of this scheme (dubbed `ISPPT') was developed and found to substantially improve the skill of weather forecasts in areas with significant convective activity \citep{Christensen2017}. In \citep{MacLeod2016a}, stochasticity was added to the land-scheme of the IFS, and was found to have a positive impact on seasonal predictability, as well as the representation of the 2003 European heatwave. 

\citep{Strommen2019} found that SPPT, the stochastic land scheme, as well as another atmospheric scheme called `independent SPPT' (\citep{Christensen2017}, substantially affected the mean state of the EC-Earth model by considering an ensemble of simulations with forced SSTs. All three schemes, particularly the two atmospheric schemes, were found to notably change the model's energy budget and thereby surface temperature. In addition, both the vertical distribution of cloud cover and the liquid water content of the clouds were found to change when turning on the stochastic schemes, and these changes were hypothesized to be responsible for the major changes in the energy budget and hydrological cycle.

In this paper we demonstrate, for the first time, that SPPT also substantially affects 21st century global warming in a coupled general circulation model, reducing it by 10\%. The scheme also changes the model's mean state, in particular reducing the global mean surface temperature. Both changes are linked to a change in the modeled clouds due to SPPT. 

The paper is structured as follows. Section \ref{sec:data_methods} contains information about the data used: we describe the EC-Earth model, the experiments considered and the SPPT scheme. Statistical methods are also described. In Section \ref{sec:results}, we document and visualize the results of the paper: we show the change in historical mean temperatures and the reduced climate sensitivity. Section \ref{sec:analysis_mean} contains our analysis on mean state changes. These are linked, to first order, to rapid changes in the cloud liquid water content of clouds through most of the vertical layers. This has the effect of increasing the optical thickness (and hence albedo) of clouds, reducing incoming solar radiation and thereby cooling the surface. The reduced level of global warming is analysed in Section \ref{sec:analysis_climsens}, and is linked to two factors. Firstly, the scheme appears to reduce the low-level cloud cover feedback by slowing down the trend of reduced low-level cloud cover seen in the model. This may be because clouds with more water tend to break up less easily when subjected to an increased temperature. Secondly, the increased cloud water due to SPPT leads to a slight increase in the negative cloud optical feedback. We also discuss the possible impact of potential non-linearity in global warming feedbacks. Concluding remarks are made in Section \ref{sec:conclusions}.

\section{Data and Methods}
\label{sec:data_methods}

\subsection{The EC-Earth Model}
\label{sec:ecearth_descr}

EC-Earth v3.1 is an Earth system model developed by the international EC-Earth consortium \citep{Hazeleger2012}. The atmospheric component uses a modified version of the Integrated Forecast System (IFS) used by the European Centre for Medium-Range Weather Forecasts (ECMWF). Land-surface processes are simulated using the Hydrology Tiled ECMWF Scheme of Surface Exchanges over Land (H-TESSEL) \citep{Balsamo2009}. The atmosphere is dynamically coupled to the ocean model `Nucleus for European Modelling of the Ocean' (NEMO) model version 3.6. The coupling in this case is handled with OASIS3 \citep{Valcke2013}. For all experiments considered, the spectral truncation (i.e. resolution) of the IFS component is T255 (corresponding to roughly 80km grid-spacing near the equator), and the ocean resolution is $1^{\circ}$.

\subsection{Description of the Stochastic Scheme SPPT}
\label{sec:sppt_descr}

The SPPT scheme has been used in ECMWF's operational ensemble forecasts since 1998, and is designed to represent forecast uncertainty that arises from the unpredictable sub-grid scale variability. This is done by perturbing the total net tendency from the physics parametrisations using multiplicative noise:
\begin{equation}
  \mathbf{P}_{perturbed} = \left( 1 + \mu r \right) \sum_{i=1}^{6} \mathbf{P}_i,
  \label{eq:SPPTpert}
\end{equation}
where $\mathbf{P}$ is the tendency vector (of a given variable) from the $i$-th physics parametrisation scheme and $r$ a random variable. Note that only tendencies for prognostic model variables (winds, temperature and specific humidity) are perturbed. Diagnostic variables (such as cloud water) are computed as normal using the prognostic variables. The perturbation $r$ is a random scaling factor that is constant in the vertical, with the scaling tapered by $\mu \in \left[0,1\right]$, which is smoothly reduced to zero in the boundary layer and stratosphere, and 1 otherwise. Roughly speaking, $\mu = 1$ between pressure levels $900hPa$ and $100hPa$ (see \citep{Palmer2009}). Furthermore, $r$ follows a Gaussian distribution with mean zero, and is smoothly correlated in space and time. The implementation in EC-Earth follows that in the Integrated Forecasting System as described in \citep{Palmer2009}. The total perturbation $r$ is generated by summing over three independent spectral patterns with standard deviations [0.52, 0.18, 0.06], spatial correlation lengths [500 km, 1000 km, 2000 km] and temporal decorrelation scales [6 hours, 3 days, 30 days] respectively. The perturbation $r$ is limited to the range zero to two to ensure $\mathbf{P}_{perturbed}$ has the same sign as $\mathbf{P}=\sum \mathbf{P}_i$, with this latter being the `deterministic' tendency.

Detailed information about the physics parameterisations and computation of diagnostic variables in the IFS can be found in \citep{ifsdoc}.

\subsection{The SPHINX Experiments}

This study was based on 6 coupled simulations of the EC-Earth model carried out as part of the `Climate SPHINX Project', hereby referred to simply as SPHINX: `Stochastic Physics and HIgh resolution eXperiments' (\citep{Davini2017}). They each span the years 1850-2100, with historical forcings before 2010 and RCP8.5 forcings (\citep{Riahi2011}) from then on. Three simulations were `deterministic' (i.e. without SPPT) and three were run with SPPT. Each set of simulations started from their own ocean spin-up. The deterministic spin-up was run with constant 1850's forcing for 100 years to generate an initial model state: tiny perturbations to this initial state were used to generate the three distinct initial conditions for the ensemble members. The initial conditions for the stochastic simulations were generated similarly, except the SPPT scheme was turned on in the initial spin-up phase. 

Further details of the model configuration can be found in \citep{Davini2017}. Of particular note is the introduction, in the stochastic version of the model, of a `humidity fix'. It was found that the SPPT scheme does not conserve water, leading to an unphysical drying of the atmosphere. The `fix' computes, at each time-step, global mean precipitation and evaporation, and reinserts the amount of humidity required to bring these into balance. This humidity is reinserted with spatial weighting favouring regions where the imbalance is large. The impact of this fix on the results are discussed in the next section.

\subsection{The `FastSPHINX' Experiments}
\label{sec:fastsphinx}

Because the deterministic and stochastic initial conditions of the SPHINX experiments had already been allowed to diverge as part of their individual spin-up phases, it is not possible to identify the fast changes in the model caused by turning on SPPT from the SPHINX simulations alone. This makes the determination of any root cause of mean state changes, which are often present at the start of the simulations proper, very difficult, given the strong coupling between multiple key variables. In particular, we will show that SPPT systematically affects clouds, and it is not possible to determine from the SPHINX experiments if the cloud changes were forcings of, or responses to, the temperature changes. Therefore, we additionally performed 10 pairs of new simulations using the exact same model as used in SPHINX. Five of these pairs were initialized in February of 1850, five in August of the same year. For each start-date, 5 simulations were run with SPPT and 5 without. Crucially, all the pairs of deterministic/stochastic simulations started from the same initial condition. In each pair, the SPPT simulation diverges from the deterministic counterpart purely due to the turning on of the stochastic scheme, and therefore these experiments allow us to identify the rapid response in the model to SPPT. We will refer to these experiments informally as the `FastSPHINX' experiments, to distinguish them from the SPHINX experiments. 

The initial state used for these experiments was the initial (spun up) state of the deterministic SPHINX simulations, and therefore are run from an ocean initialized with constant 1850's forcing. Small perturbations to this state are allowed to evolve for 1 year with this fixed forcing to produce 5 distinct initial conditions for each of the two seasons. Each deterministic/stochastic pair is then run for 6 months, again with fixed 1850's forcing; the final data analysed is comprised of these final 6 months. The choice of both February and August start-dates was made to account for possible seasonal influences on the response to SPPT. For the results shown, no meaningful such influence was found, so we will typically present data using all start-dates with no further comment.

To test the robustness of the results, additional experiments were also performed. Firstly, a similar ensemble with forced SSTs was carried out, and secondly, an experiment to test the potential impact of the `humidity fix' was also carried out by running a simulation with SPPT turned on but the `fix' turned off.

In all simulations, the tuning parameters of EC-Earth are identical for both deterministic and stochastic simulations. Consequently, the \emph{only} difference between the deterministic and stochastic model is the turning on of SPPT in the latter.

\subsection{Statistical Methods}
\label{sec:stats}

When considering the SPHINX simulations, we do not apply formal statistical tests, due to the small sample size of 3. Instead, we typically allow the data to speak for itself by plotting all ensemble members together. For the large changes observed, all three stochastic ensemble members will usually be greater/smaller than all three deterministic counterparts. For long timeseries, such a clear divergence over multiple decades would be highly unlikely if the distributions drawn from are in fact the same. In the absence of further ensemble members, a clear such divergence over time will be taken as our criteria for significance when considering SPHINX data.

For the FastSPHINX experiments, statistical significance is calculated using a two-tailed T-test, with no assumption of equal variance across the samples. When means are estimated, errorbars (e.g. indicated with shading on timeseries plots) are defined using the standard error. The standard error is computed as the standard deviation across the sample divided by the square root of the number of samples (i.e. the square root of 10 in our case). When considering differences, we always pair up the corresponding deterministic/stochastic simulations that started from the same initial condition.

\section{Results}
\label{sec:results}

\begin{figure}[p]
    \includegraphics[width=12cm]{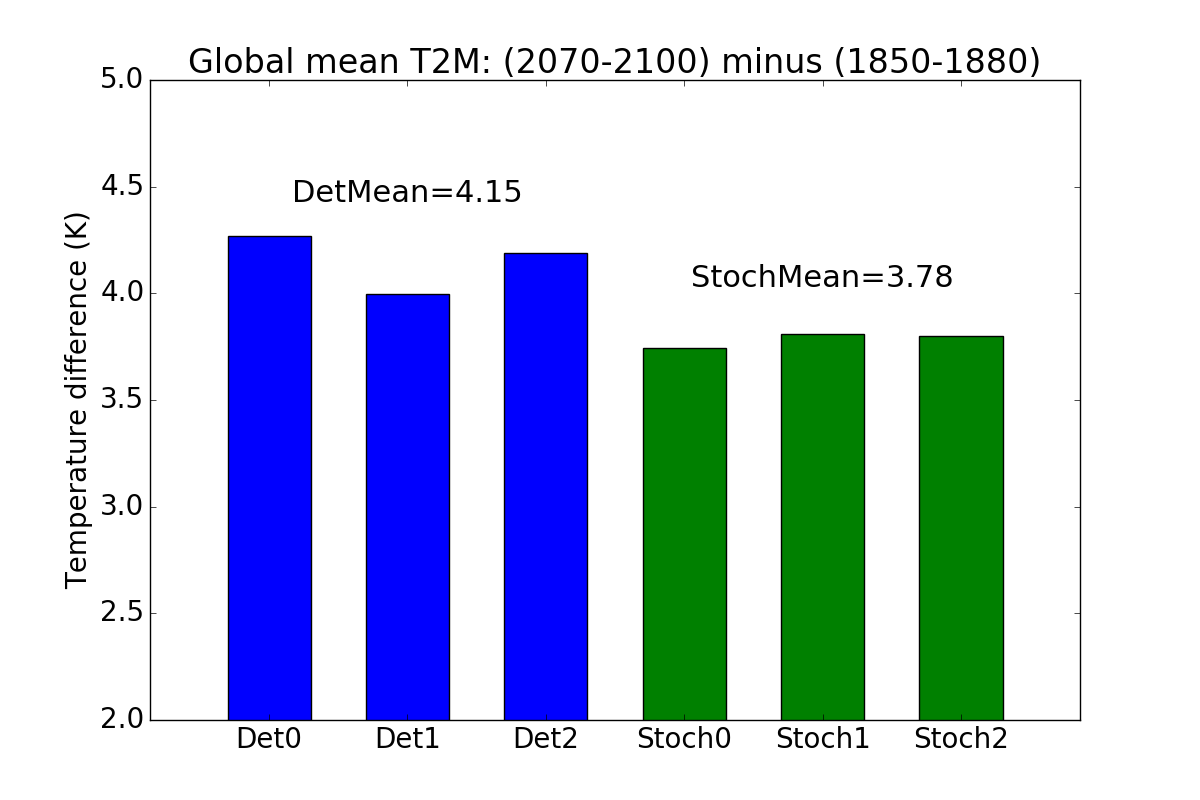}
	\caption{Extent of global waming by 2100 for three deterministic simulations (Det0, Det1 and Det2 in blue) and three stochastic simulations (Stoch0, Stoch1 and Stoch2 in green): global mean surface temperature over last 30 years of simulation (2070-2100) minus same over first 30 years (1850-1880). DetMean (respectively StochMean) denotes the mean of the three deterministic (respectively stochastic) values.}
	\label{fig:climsens}
\end{figure}

For a given SPHINX simulation, we measure the extent of (transient) global warming by taking the difference between the global mean surface temperature averaged over the first 30 years of the simulation (1850-1880) and the last 30 years (2070-2100). We will refer to this as the model's greenhouse-gas sensitivity, or GHG-sensitivity for short: it should not be confused with either the equilibrium climate sensitivity or transient climate response. Figure \ref{fig:climsens} shows the GHG-sensitivity for the three deterministic and three stochastic simulations. It can be seen that all three deterministic simulations have a higher GHG-sensitivity, at 4.15K on average, than all three stochastic simulations, at 3.78K on average. Therefore the stochastic scheme has reduced the projected global warming by about 10\%.

\begin{figure}[p]
    \includegraphics[width=12cm]{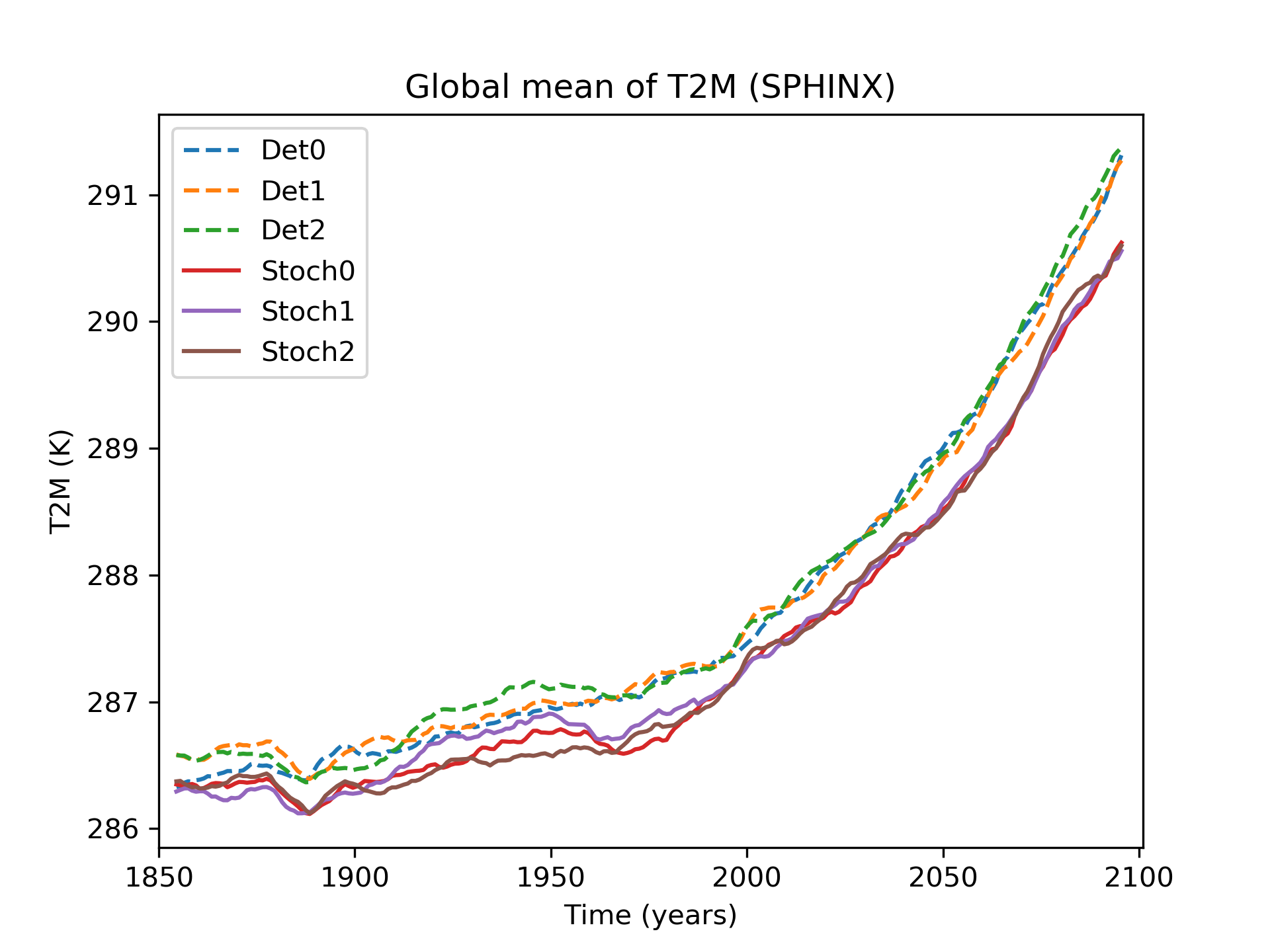}
	\caption{Evolution of global means of surface temperature (T2M) of the three deterministic simulations (Det0, Det1, Det2: solid lines) and the three stochastic simulations (Stoch0, Stoch1, Stoch2: stipled lines). A 10-year running mean has been applied to smooth the data.}
	\label{fig:temp_evolution_all}
\end{figure}

\begin{figure}[p]
    \includegraphics[width=12cm]{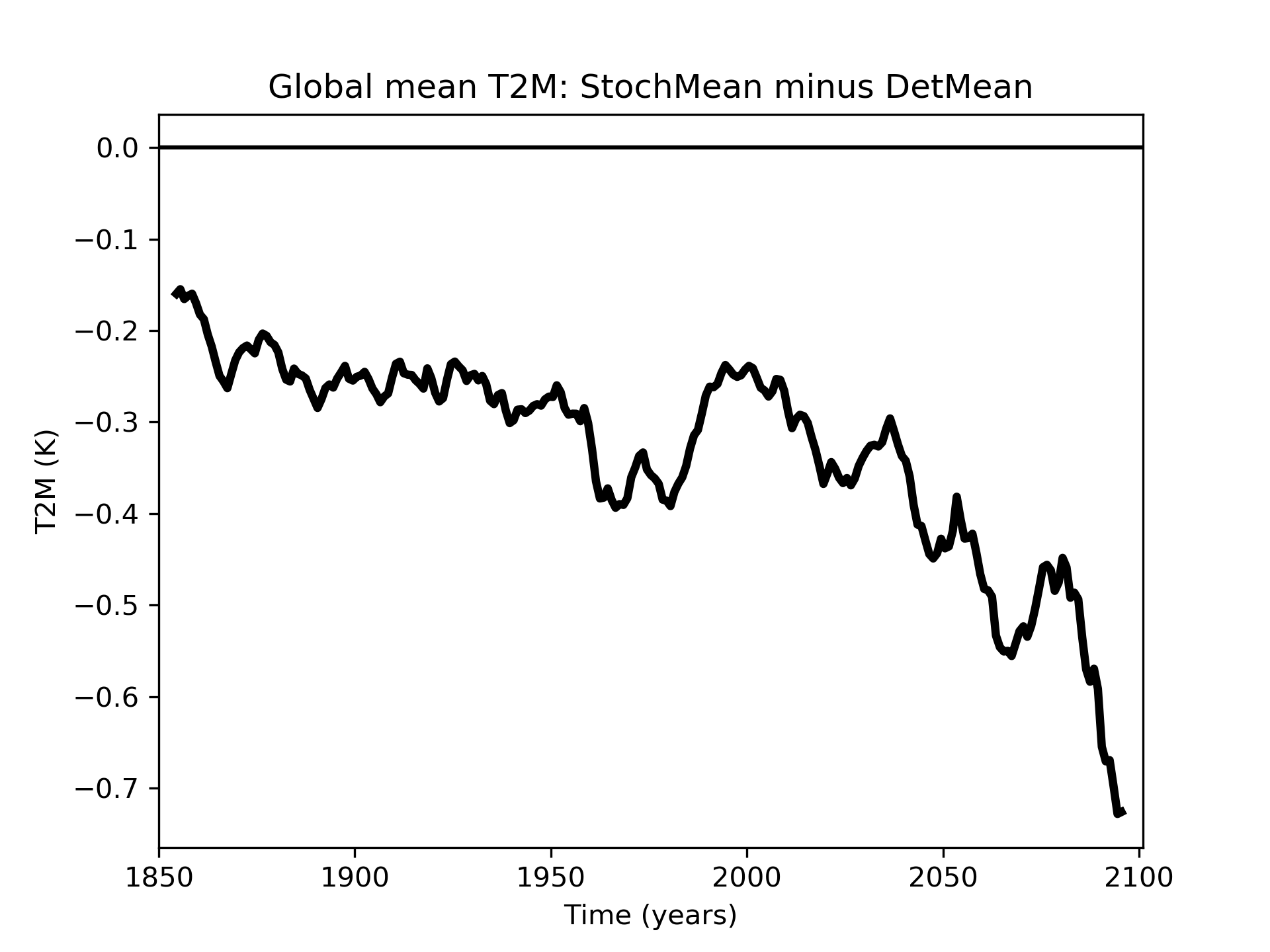}
	\caption{Timeseries of the difference in ten-year running means of ensemble-mean global mean temperature between the stochastic and deterministic simulations.}
	\label{fig:temp_evolution_diff}
\end{figure}

Figure \ref{fig:temp_evolution_all} shows the temporal evolution of global mean surface temperature for all the individual simulations. A 10-year running mean has been applied to isolate changes on climate timescales.  The deterministic and stochastic ensembles are robustly separated across the entire time period, with the stochastic simulations having an overall cooler mean climate. The increased divergence that takes place as one approaches the end of the 21st century is more apparent in Figure \ref{fig:temp_evolution_diff}, which shows the difference in the stochastic and deterministic ensemble means. During the 20th century there is a fairly constant difference in mean surface temperature of around 0.3K, which begins to grow rapidly from around 2040 onwards. While emissions plateau towards the end of the 21st century in RCP8.5, there is no obvious sign that the divergence between the models has done the same. This may be simply due to the inherent lag in the system's response to forcing, but it also raises the possibility that the ultimate difference in equilibrium climate sensitivity may be greater than the 10\% computed here.

\begin{figure}[p]
    \includegraphics[width=12cm]{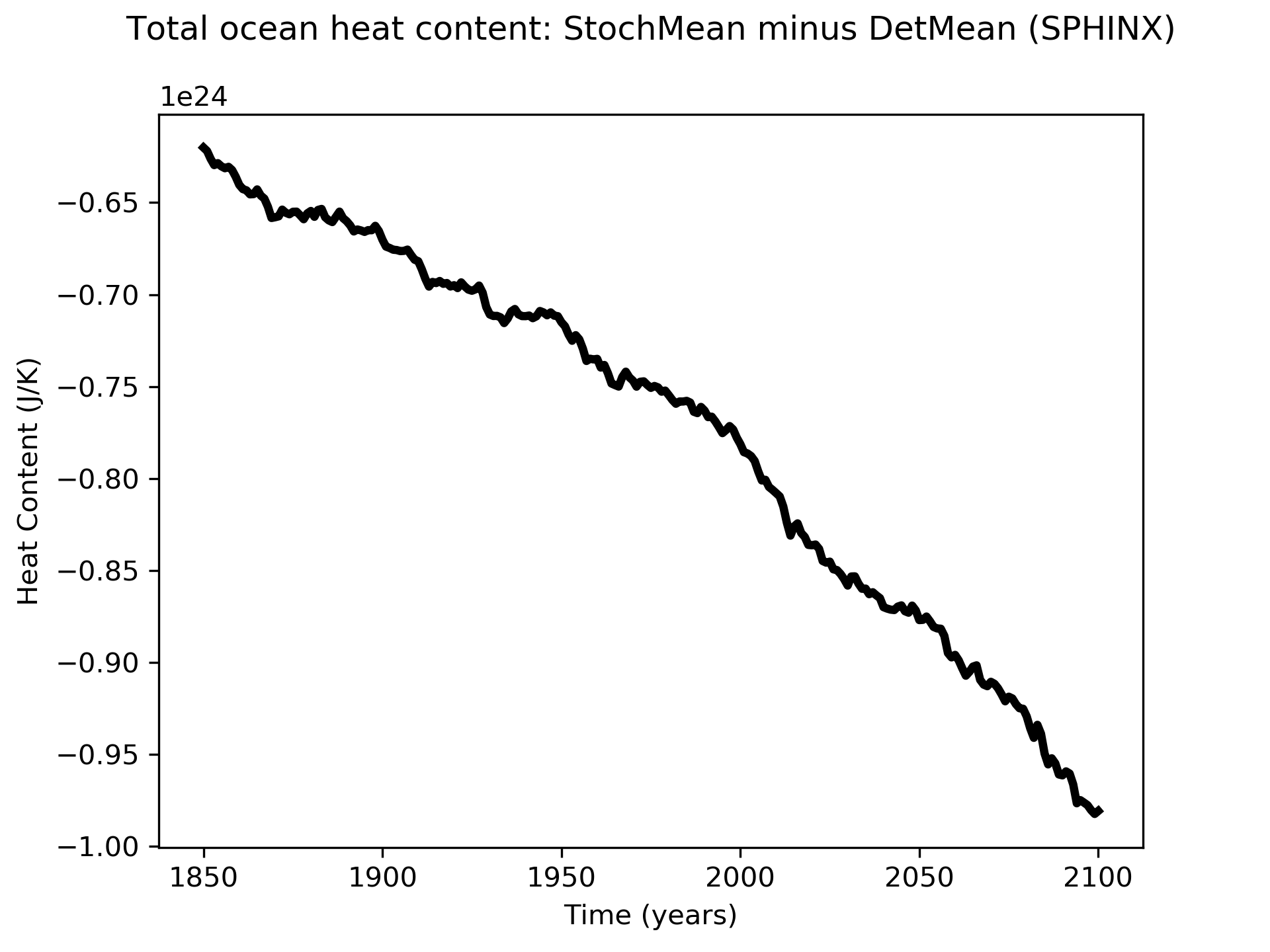}
	\caption{Timeseries of the difference in ten-year running means of ensemble-mean ocean heat content between the stochastic and deterministic simulations.}
	\label{fig:oceanheatc}
\end{figure}

Figure \ref{fig:oceanheatc} shows the difference between the stochastic and deterministic ensemble-mean ocean heat content. In both simulations, the ocean heat content is increasing in the same manner as surface temperature (not shown). The fact that the difference is becoming more negative implies that SPPT is inhibiting the warming not just of the atmosphere, but of the system as a whole. This also verifies that the results are not due to issues with the ocean spin-up.

\begin{figure}[p]
    \includegraphics[width=12cm]{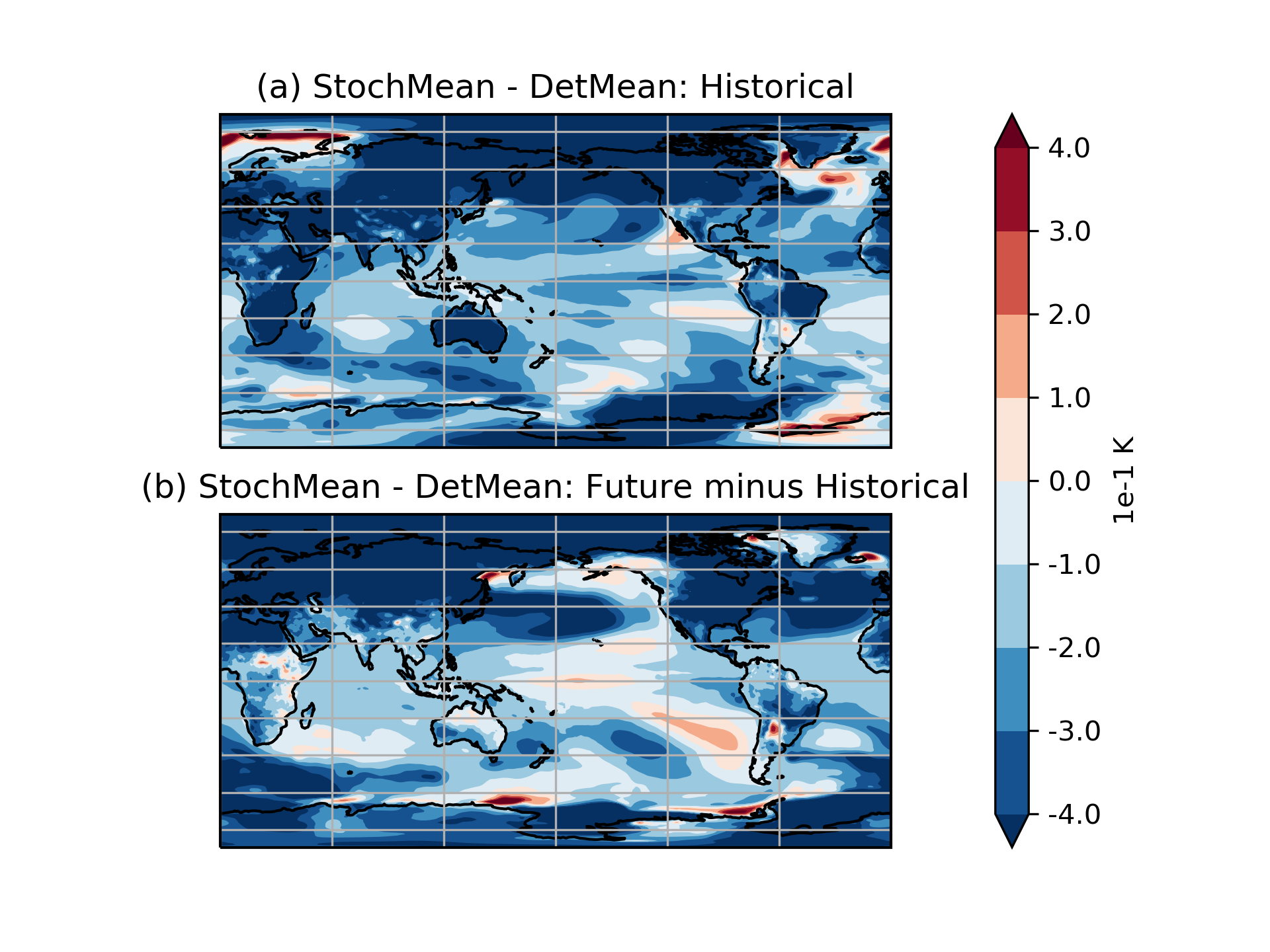}
	\caption{Evolution of mean surface temperature differences (K) between historical period and future period for the SPHINX experiments. In (a) Stochastic ensemble mean (StochMean) minus Deterministic ensemble mean (DetMean)  over the period 1980-2010. In (b), the stochastic ensemble's mean warming between 1980-2010 and 2070-2100  minus that in the deterministic ensemble}
	\label{fig:spatial_temp_evolution}
\end{figure}

Figure \ref{fig:spatial_temp_evolution} shows the spatial pattern of the temperature differences in the historical period (1970-2000), and how this difference has changed at the end (2070-2100) of the simulation. It can be seen that the bulk of cooling with stochastic physics accumulates in the northern hemisphere (NH), particularly over land and the Arctic. This tendency is amplified further by the end of the 21st century, with increased cooling relative to the deterministic model in the same regions. A possible explanation for this NH accumulation is the phenomenon of Arctic amplification of global mean surface temperature changes due to anthropogenic forcing (\citep{Pithan2014, Barnes2015}).

Through the rest of the paper we will aim to understand the drivers behind the evolution seen in Figure \ref{fig:temp_evolution_diff}, including both the initial divergence leading to the cooler model climate with SPPT in the historical period, as well as the later increased divergence towards the end of the 21st century.

\section{Analysis: Change in the Mean State}
\label{sec:analysis_mean}

Changes in global mean surface temperature (T2M) are associated with imbalances in the energy budget, since surface temperatures will change in order to bring the system to equilibrium. Therefore we will take as our starting point the model's surface energy budget, and use this to determine the dominant drivers of change when turning on SPPT.

Figure \ref{fig:energy_full} shows the differences in the fluxes making up this energy budget between the stochastic and deterministic ensemble means. Note that we have used the convention that downward fluxes are positive. In particular, an increase in downward latent heat flux due to SPPT corresponds to a \emph{decrease} in evaporation (an upward flux), and vice versa. The upshot of this convention is that the net surface energy can be conveniently obtained by adding up the other fluxes, and so the dominant source of change can be identified at any point in time. As before, we have smoothed the timeseries by a 10-year running mean, in order to retain only the fluctuations taking place on climate timescales.

During the 20th century the mean difference in the net flux is around $-0.1 W/m^2$, which grows to around twice that by the end of the 21st century, consistent with both the cooler mean climate and the reduced  GHG-sensitivity. Anthropogenic forcings lead to a net positive surface energy imbalance for both configurations, which both sets the historical mean temperature, and, as this imbalance increases throughout the RCP8.5 scenario, the extent of global warming. The reduction of this excess amount of energy due to SPPT in the historical period leads to a cooler historical climate, and the growth of this reduction implies a weaker GHG-sensitivity.

We will begin by trying to understand the initial lower average net surface energy flux, and then follow up by trying to understand the decreased GHG-sensitivity.

\subsection{Historical Energy Budget Analysis}
\label{sec:analysis_historical}

Figure \ref{fig:energy_full} shows that the dominant contribution to the decreased surface energy flux due to SPPT is a strong decrease in surface solar radiation (SSR). The second consistent, but lesser, contribution, is made by a decrease in surface thermal radiation (STR), associated with a decrease in the amount of longwave radiation emitted to the surface by the atmosphere. This is likely due to a decrease in the global mean atmospheric water vapor content due to SPPT (not shown), which reduces the greenhouse effect. However, as will be seen in the next section, this change appears to be a \emph{response} to the cooling induced by reduced solar radiation, so we do not discuss this further.

Changes in net solar radiation will be determined by changes to the albedo of the surface, the cloud coverage and the albedo of the clouds. This latter property, a function of the optical thickness of the cloud, depends to a large extent on the cloud liquid water content (CLWC) of the clouds, i.e. the total amount of liquid water per unit volume of air in a cloud parcel; this is sometimes also referred to as the `liquid water path' (\citep{Han1998}). Changes in surface albedo are on the other hand going to be associated with changes in snow and ice cover; EC-Earth does not have dynamic vegetation, so the albedo of the land will not otherwise change. By decomposing the net SSR into a contribution from the central region 60S-60N and a high-latitude region (the combined regions 60-90N and 60-90S), one finds that ${\sim}  86\%$ of the difference in SSR due to SPPT can be accounted for by differences in 60S-60N, where surface albedo changes are small. This suggests that changes in sea-ice and snow coverage are not primarily responsible. Computing timeseries of global mean sea-ice volume and snow coverage supports this conclusion, showing no notable differences between the stochastic and deterministic simulations during the entire historical period (not shown).

\begin{figure}[p]
    \includegraphics[width=12cm]{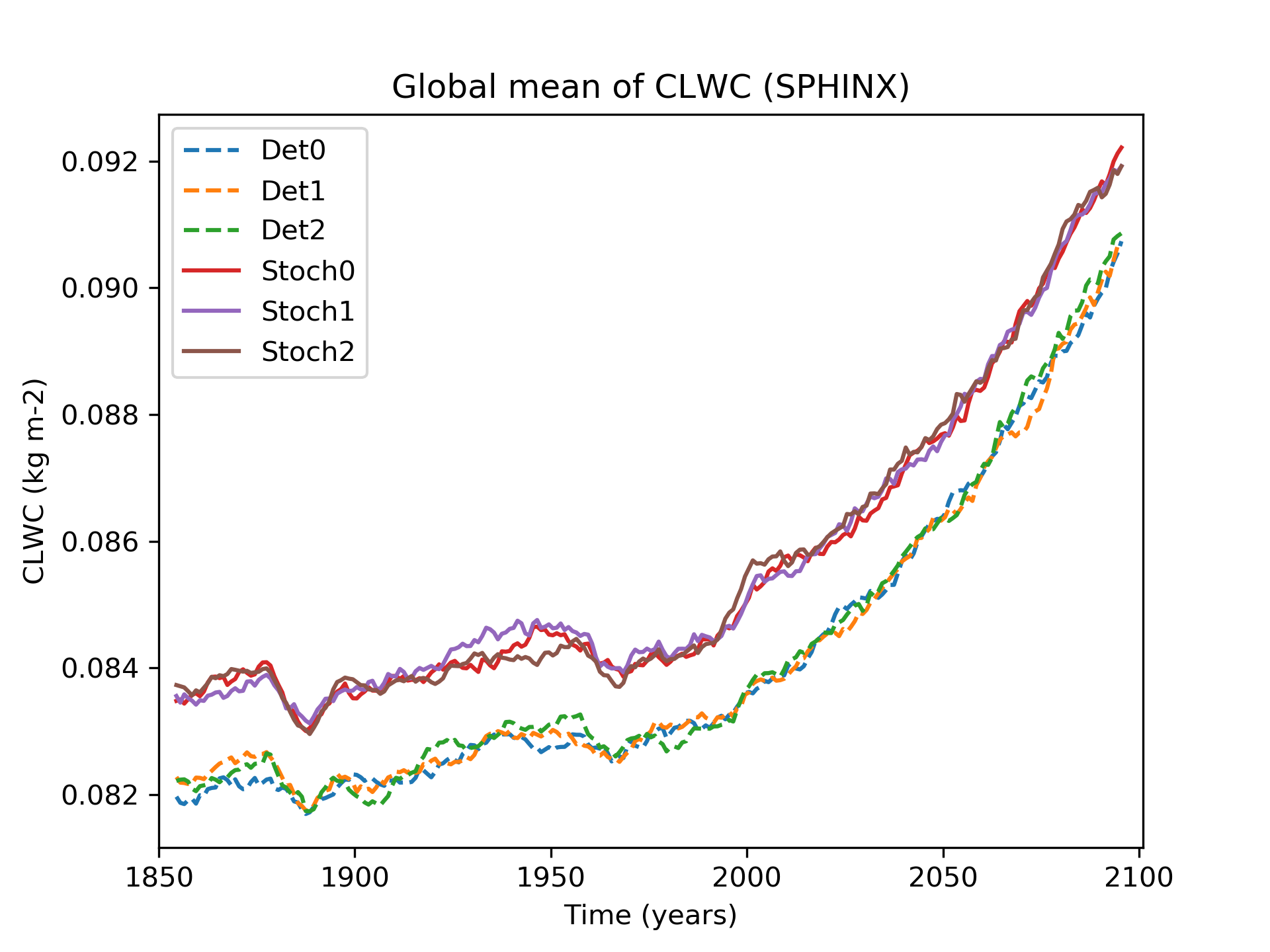}
	\caption{Evolution of vertically integrated cloud liquid water content (CLWC) for the SPHINX simulations. A 10-year running mean has been applied to smooth the timeseries.}
	\label{fig:CLWVI_timeseries}
\end{figure}

\begin{figure}[p]
    \includegraphics[width=12cm]{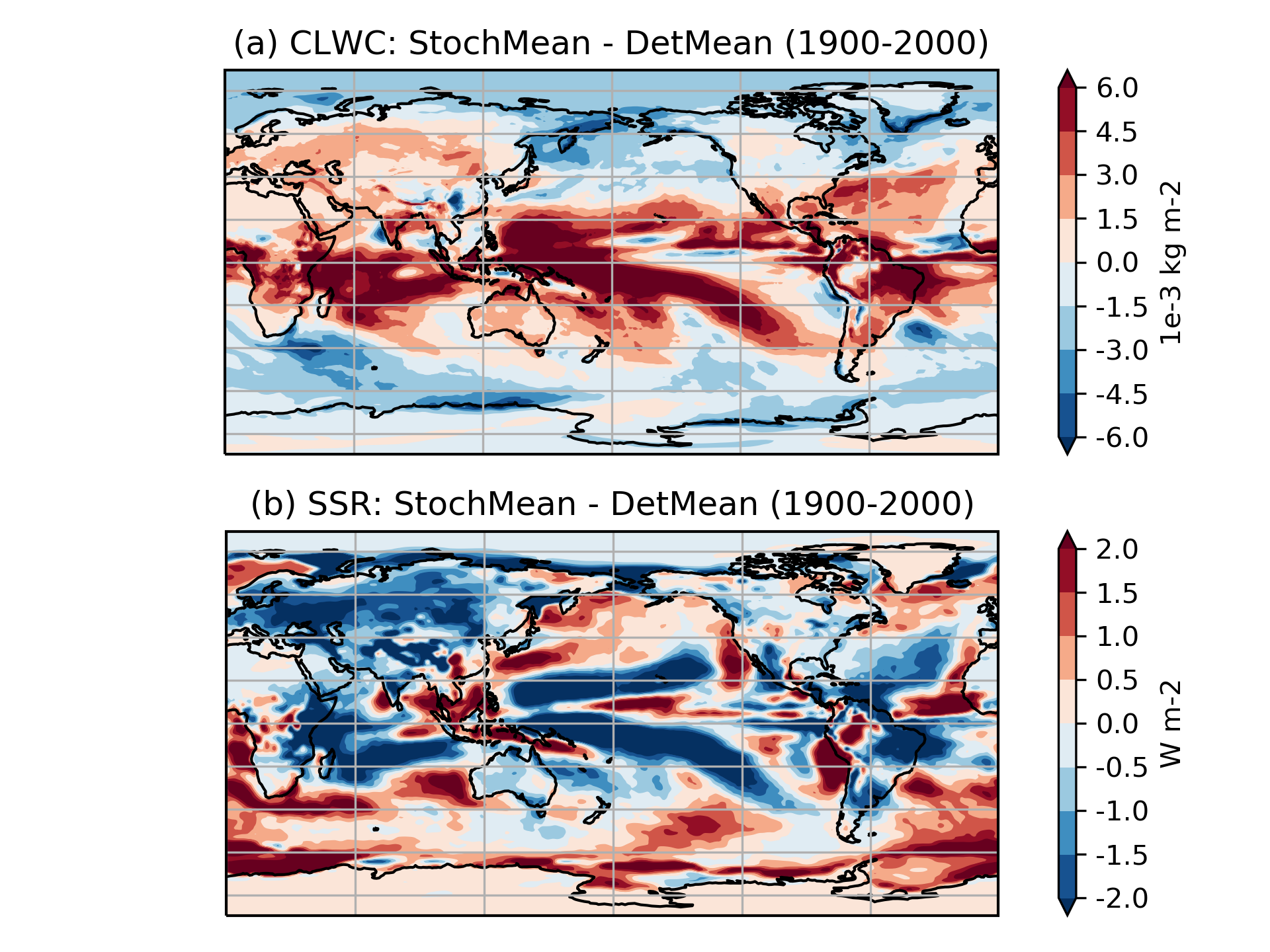}
	\caption{Stochastic minus deterministic ensemble mean (1900-2000) for (a) cloud liquid water content (CLWC) and (b) surface solar radiation (SSR). SPHINX experiments.}
	\label{fig:clwc_ssr_sphinx}
\end{figure}

Turning then to changes in cloud coverage, Figure \ref{fig:tcc_diffs_sphinx} shows that between 1850 and 2000 there is no notable change in the total cloud cover (TCC) between the two ensembles. On the other hand, Figure \ref{fig:CLWVI_timeseries} shows a robust increase in CLWC with SPPT across the entire simulation period. The difference is approximately constant in time and represents an increase of about 1.6\% relative to the deterministic mean; this is consistent with the findings of \citep{Strommen2019}, who find a similar increase in CLWC in a different version of EC-Earth. Figures \ref{fig:clwc_ssr_sphinx}(a) and (b) show the spatial structure of CLWC and SSR changes respectively. The majority of the CLWC change is concentrated in the tropical Pacific and the Indian Ocean. In these regions there is good spatial coherence between CLWC and solar radiation changes, with a pattern correlation between the two fields, when restricted to 60S-60N, of around -0.63. It is known that the dependence of cloud albedo on CLWC is non-linear, with the impact being stronger for optically thin clouds than for optically thick clouds (e.g. \citep{Han1998}), so correlations close to -1 would not necessarily be expected. Furthermore, while the global mean cloud cover does not change with SPPT, there are various regional changes (not shown) which may further influence the exact location of solar radiation changes. We conclude that it is likely that the increased CLWC is a major cause of the decreased solar radiation. 

To assess the model's global mean CLWC, we compared against observational data obtained via remote sensing techniques (\citep{REMS}). The data covered the period 2000-2008 and was restricted to ocean points only; by restricting the model data similarly, we found that the deterministic model has a positive bias (too much cloud water) of around 10\%. While the stochastic models have increased this bias, the effect is small relative to this initial bias. It is also important to note that the deterministic model was tuned to achieve a realistic energy budget for the historical period, which may have helped to adjust the CLWC towards a more realistic value. So it may be the case that the stochastic model would have a similar or smaller bias if it were also tuned.

The increased CLWC is present already at the start of the SPHINX simulations, implying that this change was stably in place after the model's spin-up phase. Therefore we will turn to the FastSPHINX experiments to assess this change further. In particular, we want to determine if CLWC changes are a response to some other rapid mean-state change. 

\subsection{Determination of Fast Changes due to SPPT}
\label{sec:rapid_changes}

\begin{figure}[p]
    \includegraphics[width=12cm]{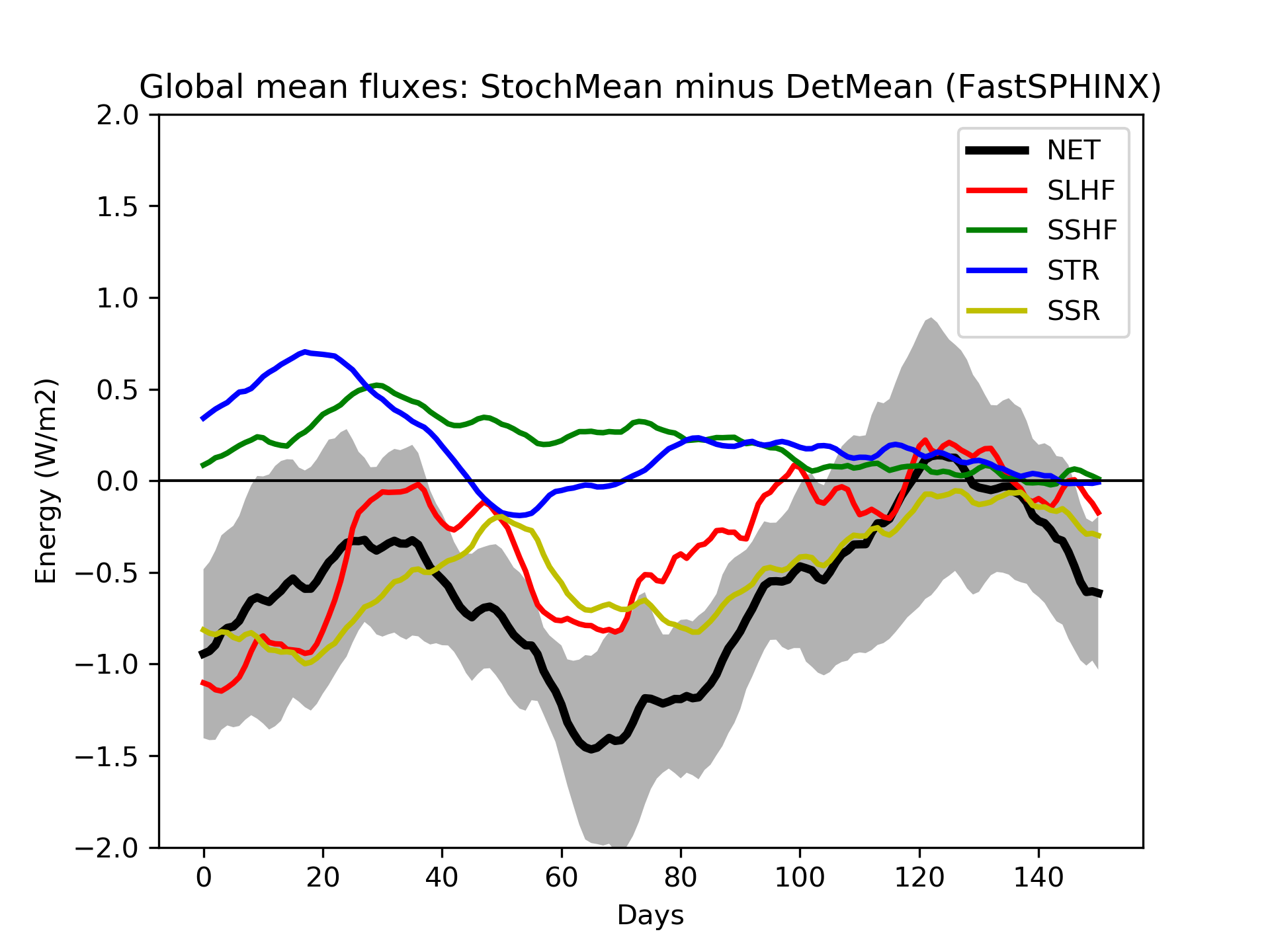}
	\caption{Evolution of global mean surface energy fluxes for the FastSPHINX experiments: stochastic ensemble mean (StochMean) minus deterministic ensemble mean (DetMean). Latent heat flux (SLHF: red), sensible heat flux (SSHF: green), thermal radiation (STR: blue), solar radiation (SSR: yellow) and net surface energy (SRF: black). The shading indicates the standard error of the mean estimate at every timepoint. A 30-day running mean has been applied to smooth all timeseries.}
	\label{fig:energy_diffs_newsphinx}
\end{figure}

As explained in Section \ref{sec:fastsphinx}, the FastSPHINX experiments will allow us to identify the rapid response to turning on SPPT. Figure \ref{fig:energy_diffs_newsphinx} shows the rapid change in surface energy fluxes: a 30-day running mean has been applied to reduce noise and highlight the more systematic trends. It can be seen that SPPT has decreased SRF, with the mean difference over the 6-month period being $-0.6 W/m^2$ . This is larger than the $0.1-0.2 W/m^2$ seen in the long climate experiments, but this is likely just reflecting the fact that it takes the model many years to reach its new equilibrium when SPPT is turned on. The dominant contribution to this decrease is a reduction in both net solar radiation and latent heat flux, with the latter (due to the sign convention) corresponding to an \emph{increase} in evaporation. Surface temperatures rapidly cool as a result of this drop in surface energy, before temporarily warming again as the system begins to respond (not shown). As we know from the SPHINX simulations, eventually a new, cooler equilibrium is reached.

\begin{figure}[p]
    \includegraphics[width=12cm]{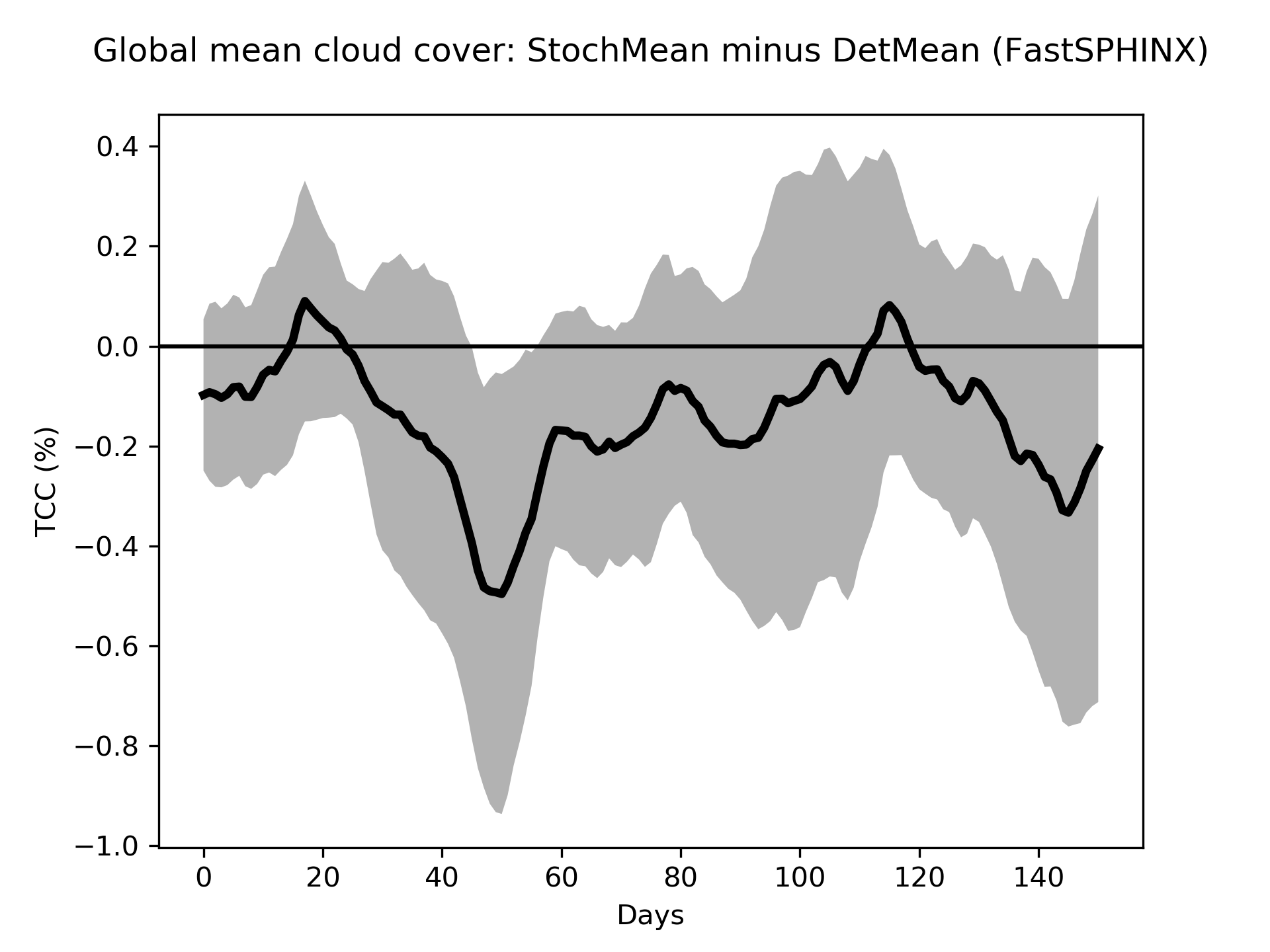}
	\caption{Evolution of total cloud cover (TCC) for the FastSPHINX experiments: stochastic ensemble mean (StochMean) minus deterministic ensemble mean (DetMean). A 30-day running mean has been applied to smooth the timeseries. The shading indicates the standard error of the mean estimate at every timepoint.}
	\label{fig:tcc_diffs_newsphinx}
\end{figure}

\begin{figure}[p]
    \includegraphics[width=12cm]{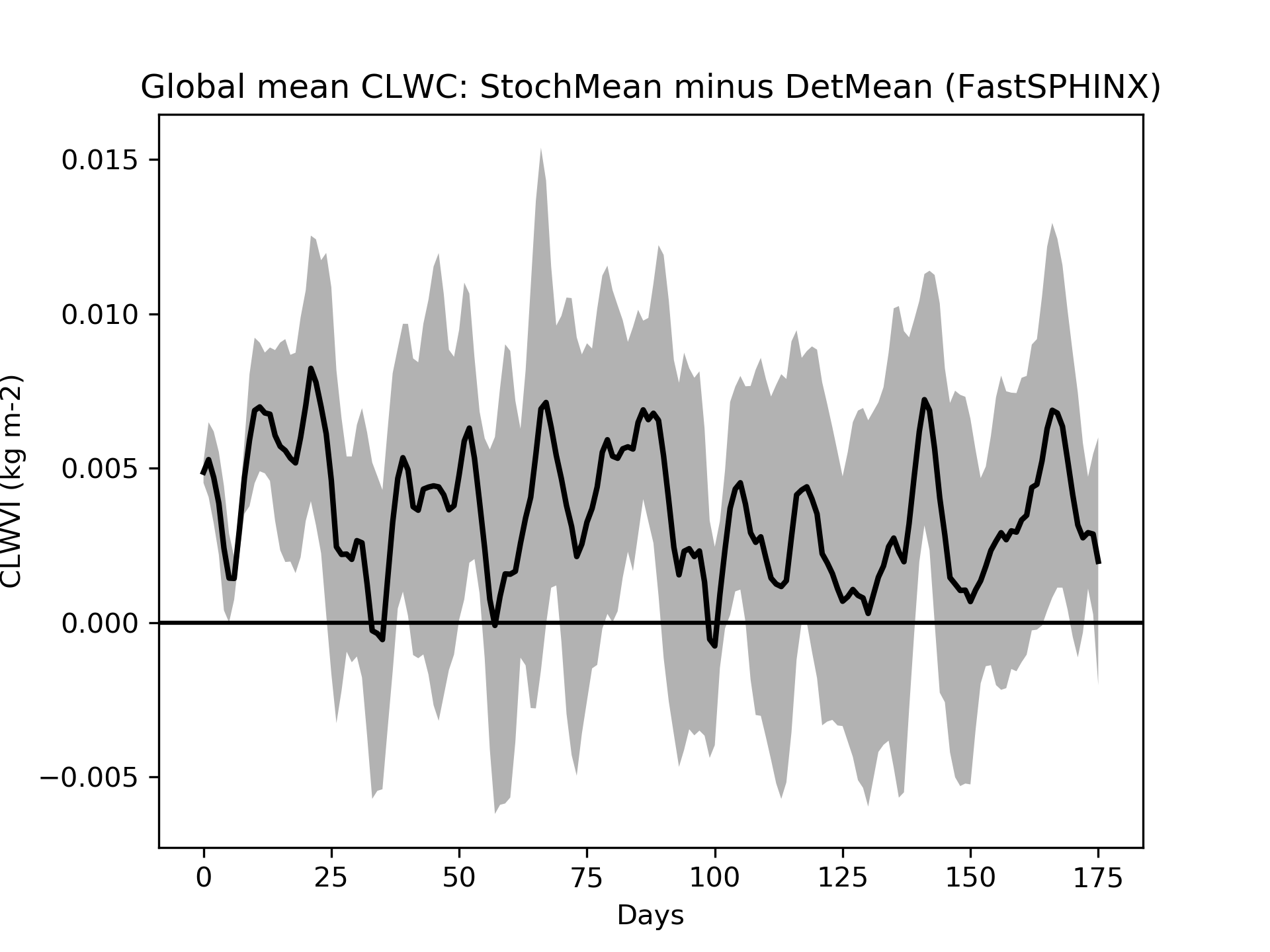}
	\caption{Evolution of vertically integrated cloud liquid water content (CLWC) for the FastSPHINX experiments: stochastic ensemble mean (StochMean) minus deterministic ensemble mean (DetMean). A 5-day running mean has been applied to smooth the timeseries. The shading indicates the standard error of the mean estimate at every timepoint}
	\label{fig:clwvi_diffs_newsphinx}
\end{figure}

\begin{figure}[p]
    \includegraphics[width=12cm]{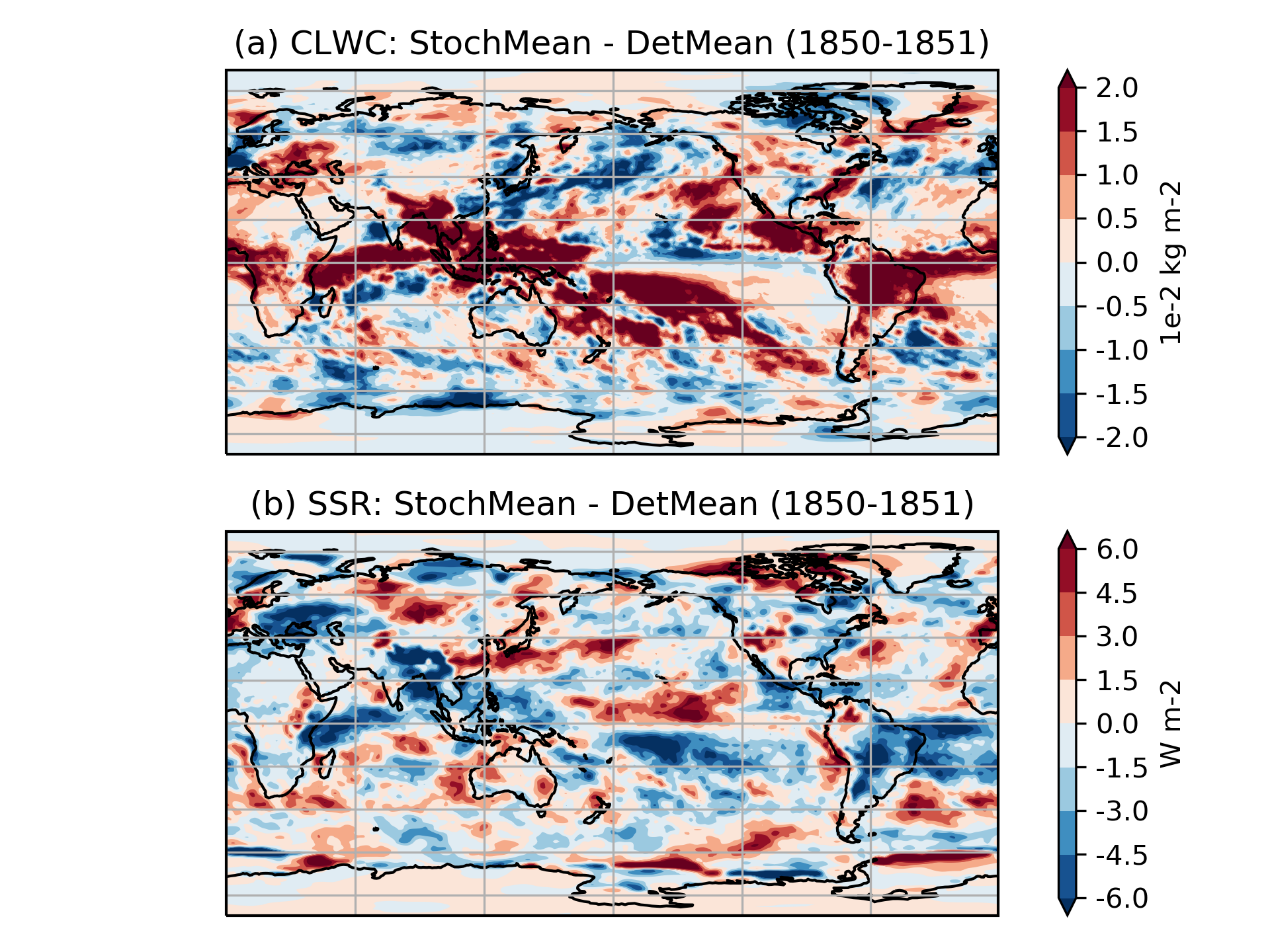}
	\caption{Stochastic minus deterministic ensemble mean (1850-1851) for (a) cloud liquid water content (CLWC) and (b) surface solar radiation (SSR), in the FastSPHINX experiments.}
	\label{fig:clwc_ssr_fastsphinx}
\end{figure}

\begin{figure}[p]
    \includegraphics[width=12cm]{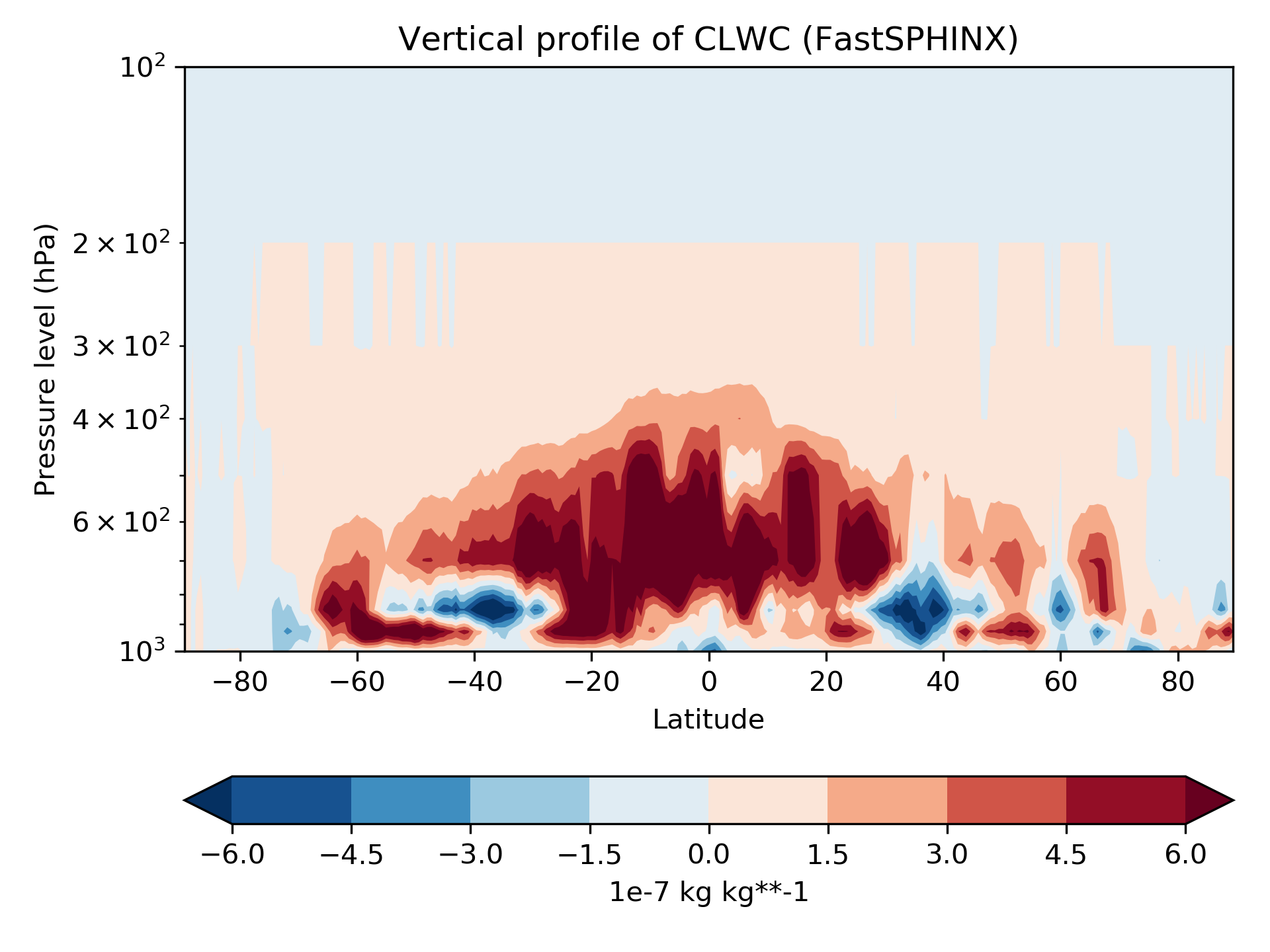}
	\caption{Cloud liquid water content (CLWC) as a function of latitude and pressure for the FastSPHINX experiments: stochastic ensemble mean (StochMean) minus deterministic ensemble mean (DetMean). Note the logarithmic scale of the y-axis.}
	\label{fig:clwvi_vertical_newsphinx}
\end{figure}

Let us first discuss solar radiation changes. Figure \ref{fig:tcc_diffs_newsphinx} shows the rapid changes in total cloud cover, showing a slight trend for the stochastic model to have reduced cloud-cover compared to the deterministic model. Because such a decrease would be expected to lead to an increase in net solar radiation, we conclude that, as with the SPHINX simulations, changes in cloud cover cannot explain the reduced solar radiation. Figure \ref{fig:clwvi_diffs_newsphinx} shows the evolution of CLWC differences. A coarser, 5-day running mean has been applied to highlight the speed of the change: the increase with SPPT, of around 2.5\%, is robust and in place within days. In fact, we find that the deterministic and stochastic simulations are statistically significantly separated in terms of their global mean CLWC within the first 6 hours (not shown). The spatial pattern of the changes are similar to those seen in the SPHINX data, as seen in Figure \ref{fig:clwc_ssr_fastsphinx}, which shows spatial changes averaged over the entire 6-month period for CLWC and SSR. As in SPHINX, the SSR changes correlate well with the CLWC changes, with a pattern correlation of -0.67 between 60S and 60N. Cloud water on individual pressure levels was not available for the SPHINX experiments, but was output from the FastSPHINX simulations. Figure \ref{fig:clwvi_vertical_newsphinx} shows the vertical structure of the changes, as a function of latitude and pressure. The increase is greatest between 400-800hPa and is less robust near the surface. This may be related to the fact that the stochastic perturbations are tapered to zero as one approaches the surface. From about 800hPa and upwards the increase is firmly in place and, whilst greatest in the tropics, extends across the full latitudinal range.

\begin{figure}[p]
    \includegraphics[width=12cm]{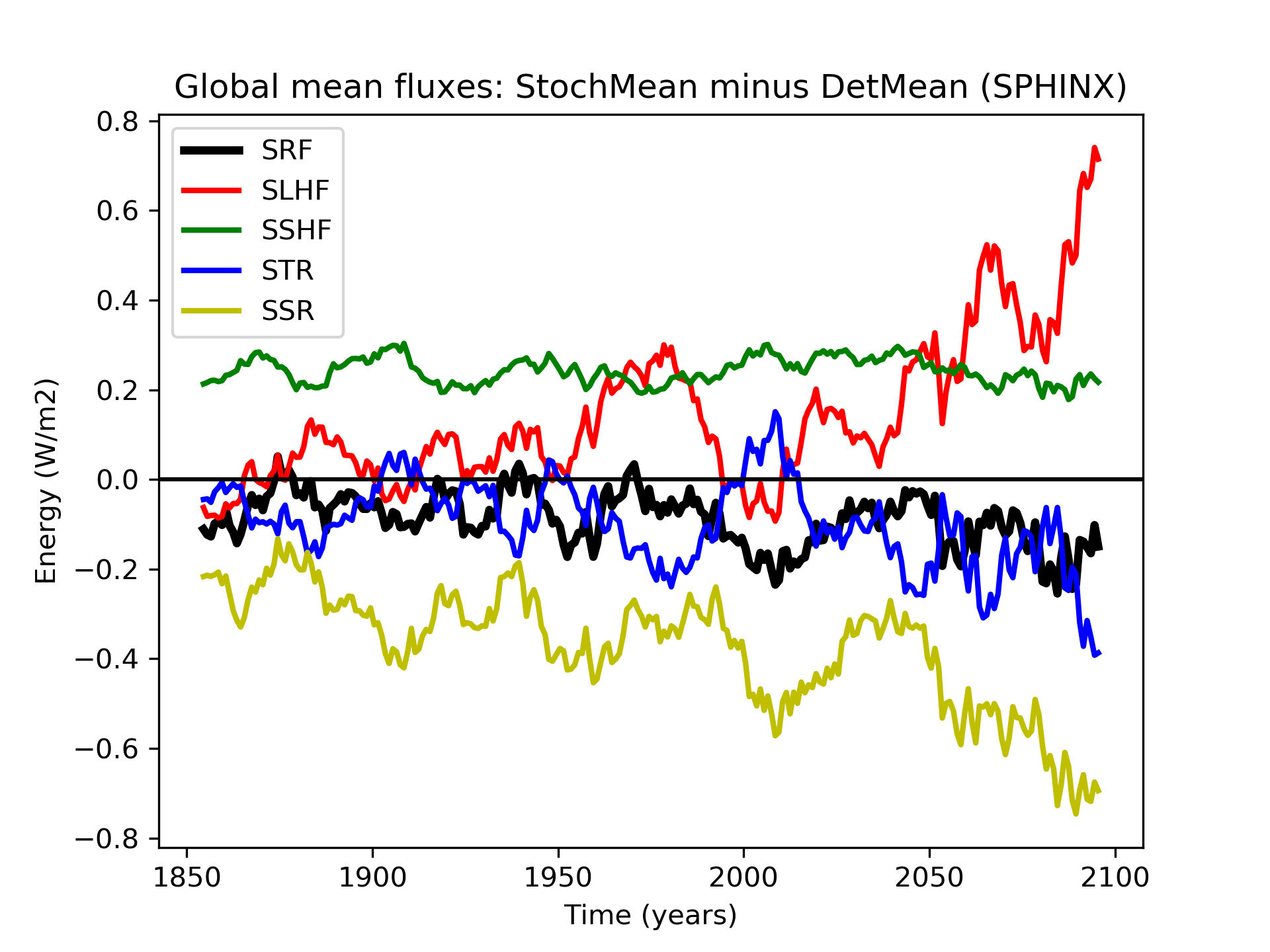}
	\caption{Evolution of surface energy fluxes ($W/m^2$): stochastic ensemble mean (StochMean) minus deterministic ensemble mean (DetMean). Surface thermal radiation (STR: blue), surface solar radiation (SSR: yellow), downward surface latent heat flux (SLHF: red), downward surface sensible heat flux (SSHF: green), net downards surface energy (SRF: black). Note we have used the EC-Earth convention that downward fluxes are positive. Thus SRF is the sum of other fluxes shown.}
	\label{fig:energy_full}
\end{figure}

Next, we consider the other major contributor to the reduced surface flux, namely the increased evaporation with SPPT. In the IFS, the amount of evaporation at a gridpoint depends primarily on the surface wind-speeds and the extent to which the specific humidity at the surface gridpoint differs from the saturation humidity (a function of surface temperature). There is also a contribution from turbulent processes, which depend on surface roughness (see \citep{ifsdoc}), but because surface roughness does not differ between the deterministic and stochastic simulations, this effect can be safely ignored. We may therefore assess the evaporation changes expected due to changes in wind speeds with fixed humidity gradient and those expected due to changes in humidity gradient with fixed wind speeds. One finds in this way that the dominant source of the increased evaporation with SPPT is due to a small, but consistent increase in surface wind-speeds (not shown). This increase in wind-speeds is also a rapid response, coming into effect within ${\sim} 1$ day. However, evaporation changes do not lead changes in cloud water, and, after a sufficiently long spin-up, become negligible (see the data Figure \ref{fig:energy_full} before ${\sim}1950$). This suggests that the impact of increased wind speeds are eventually cancelled out by the surface cooling, which reduces the saturation humidity and, therefore, evaporation.

Finally, note that the trend in surface thermal radiation is, if anything, towards a slight increase with SPPT. This implies that the reduction in this quantity seen in Figure \ref{fig:energy_full} is a long-term response to other forcings, rather than a driver of such changes. Only solar radiation changes persist from the very start of the FastSPHINX experiments through the spin-up and the subsequent SPHINX experiments. 

These results were found to be robust across both sensitivity experiments performed (described in Section \ref{sec:fastsphinx}). In the forced SST simulations, the changes in the energy budget were consistent with those seen in the coupled experiments. In particular, both sets of simulations showed the same increase in cloud liquid water. In the simulation with SPPT turned on, but the `humidity fix' turned off, the rapid response of the model is the same as that seen with the `fix' turned on, namely a sharp increase in cloud liquid water. However, the eventual drying of the atmosphere which takes place without the `fix' is, by magnitude, a much bigger change than this increased cloud water. The reduced availability of vapor for condensation thus eventually dominates, ultimately reducing the global mean cloud water content. Because the ratio of total column cloud water to total column water vapor remains approximately the same whether the `fix' is on or not, we conclude that the `fix' is not likely to be responsible for the differences found between the deterministic model and the version with SPPT including the 'fix'.

In conclusion, we find that the dominant impact to the models energy budget when turning on SPPT is to decrease the net surface solar radiation by way of a rapid increase in cloud liquid water. This increases cloud albedo, thereby reducing incoming solar radiation and, consequently, cooling the surface. The CLWC increase may be due to the stochastic perturbations interacting with the highly non-linear process of condensation. Given a parcel of air close to saturation, a perturbation of temperature or humidity in one direction may cause the parcel to condense its water, increasing the total CLWC. However, a perturbation in the opposite direction would in this situation result in no change in total CLWC. An important point here is that the implementation of SPPT used (\citep{Palmer2009}) has a `supersaturation limiter' in place, which prevents this non-linear effect from taking place within a single time-step: if the stochastic perturbations leave a gridpoint in a supersaturated state, the limiter ensures that these perturbations are not applied. Nevertheless, if the SPPT scheme pushes a parcel close to saturation, the actual model dynamics may, on the subsequent timestep, trigger condensation. In general, by broadening the distribution of humidity tendencies, one may expect to see more condensation triggered on average. Therefore, given a systematic application of symmetric, mean-zero perturbations to the tendencies, as is done with SPPT, a change in the mean CLWC is plausible.

\section{Analysis: Change in Greenhouse-gas Sensitivity}
\label{sec:analysis_climsens}

In the above analysis we linked the changes in the mean-state to rapid changes in the model's cloud properties, in particular the mean cloud liquid water content. In this section we aim to explore possible mechanisms that might explain the reduction in GHG-sensitivity. 

Figure \ref{fig:energy_full} shows that the decrease in solar radiation due to SPPT increases as the simulations progress, and remains by far the dominant source of decreased net surface energy. Given that the dominant component of spread in climate sensitivity estimates across GCMs is due to differences in cloud feedbacks (\citep{Dufresne2008}), it is plausible that these may explain the reduced GHG-sensitivity with SPPT. Cloud feedbacks can be decomposed into a contribution from three critical cloud properties: cloud altitude, cloud coverage and cloud optical depth (\citep{Zelinka2012a, Ceppi2017}). The primary impact on cloud altitude changes are on longwave radiation, while cloud cover and cloud optical depth changes are the primary sources of shortwave changes. Because changes in the shortwave radiation dominate the energy budget, we will focus on cloud cover and cloud optical depth changes: the latter is, as discussed previously, going to be largely due changes in CLWC.  


We now discuss the two key cloud feedbacks in turn, as well as the possible impact of non-linearity in climate sensitivity feedbacks.

\subsection{Differences in Cloud Cover Feedback}

\begin{figure}[p]
    \includegraphics[width=12cm]{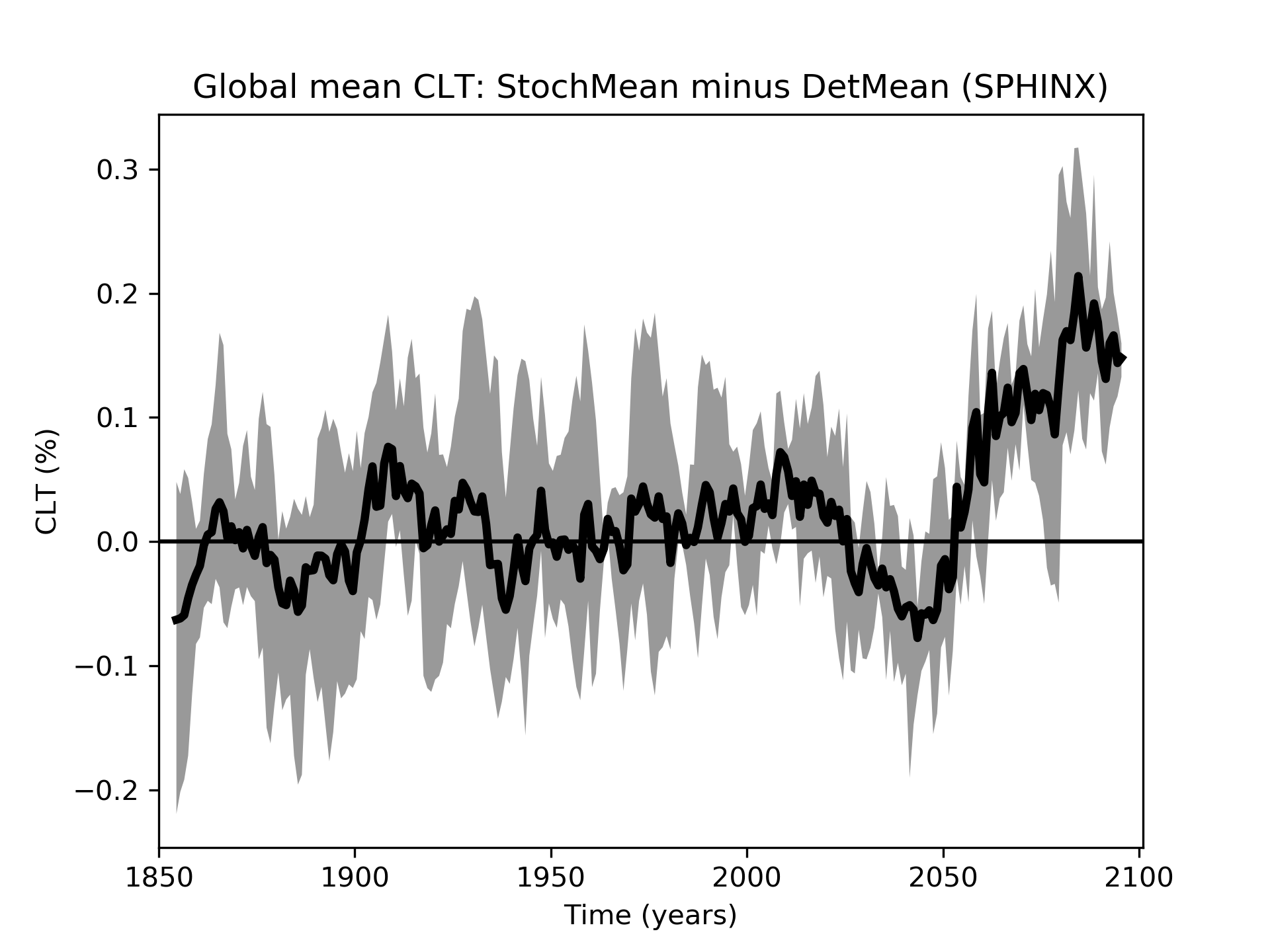}
	\caption{Evolution of total cloud cover (TCC) for the SPHINX experiments: stochastic ensemble mean (StochMean) minus deterministic ensemble mean (DetMean). A 10-year running mean has been applied to smooth the timeseries. Shading indicates the maximum/minimum difference attained across all three pairs.}
	\label{fig:tcc_diffs_sphinx}
\end{figure}

\begin{figure}[p]
    \includegraphics[width=12cm]{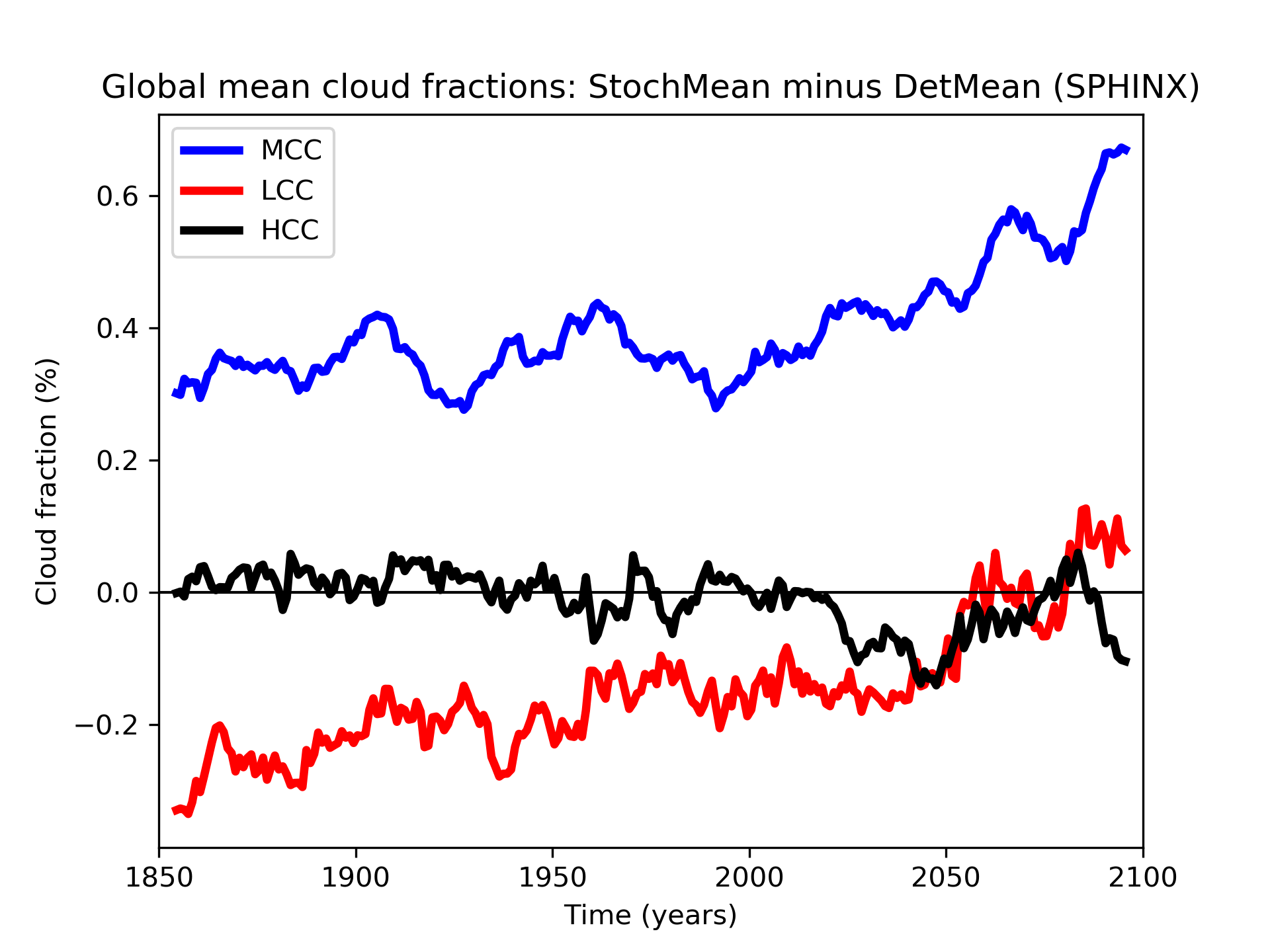}
	\caption{Evolution of cloud cover in the SPHINX simulations (1850-2100): stochastic ensemble mean (StochMean) minus deterministic ensemble mean (DetMean) for low-level cloud cover (LCC), mid-level cloud cover (MCC) and high-level cloud cover (HCC).}
	\label{fig:sphinx_clouds}
\end{figure}

Figure \ref{fig:tcc_diffs_sphinx} shows that while the stochastic and deterministic simulations do not differ notably in total cloud coverage (TCC) up to around 2050, after this date there is a robust change. For all simulations, deterministic and stochastic alike, the total cloud cover drastically reduces as global warming intensifies (not shown). Figure \ref{fig:tcc_diffs_sphinx} shows that the difference between the stochastic and deterministic simulations is becoming more positive, implying that the cloud cover is decreasing less rapidly with SPPT turned on. Figure \ref{fig:sphinx_clouds} shows again the difference in global mean cloud cover, but stratified now into low, medium and high-level clouds. SPPT has reduced low cloud cover and increased medium cloud cover, thereby effectively raising the average cloud altitude. We also see from this figure that the less rapid decline in TCC with SPPT is due to a less rapid decline in low and mid-level clouds. All three cloud types reduce as the model warms.

It is not straightforward to diagnose the cause of this reduced rate of low cloud breakup. While GCMs frequently show a decrease in low cloud cover with global warming (e.g. \citep{Zelinka2012, Zelinka2012a}), the causes for this are not well understood. \citep{Bretherton2014} discussed one potential explanation (applicable to cumulus-under-stratocumulus boundary layers and stratocumulus topped mixed layers) which they dubbed the `entrainment-liquid-flux adjustment' (ELF) mechanism. In this framework, the immediate response of a cloud to an instantaneous warming is to increase the liquid water flux everywhere in the cloud. This triggers an increased buoyancy flux, hence increased cloud-layer turbulence, which has the effect of thinning the cloud through the entrainment of warm, dry air through the cloud. With SPPT turned on, the clouds have a larger amount of liquid water in them to start with, and it is possible that this could reduce the relative impact of such an ELF response. Both theory and LES studies generally support the role of cloud water in maintaining stratocumulus clouds, and how the depletion of moisture plays a role in both the thinning of these clouds and their transition to scattered cumulus clouds with sparser coverage (\citep{Wood2012, Sandu2011}). Again, the greater availability of cloud water in the stochastic simulations may slow down such thinning and/or stratocumulus-to-cumulus transitions.

\subsection{Differences in Cloud Optical Feedback}

The cloud optical feedback states that an increase in CLWC under warming would increase cloud albedo, a cooling effect. CLWC may be expected to increase under global warming for three reasons. Firstly, if one assumes that the amount of cloud water within a given cloud follows a moist adiabat, then one can show that CLWC always increases with temperature (\citep{Betts1987, Ceppi2017}). Secondly, the increased humidity expected with warmer surface temperatures (due to the increased saturation mixing ratio) implies a greater availability of water vapor for condensation (\citep{Somerville1985}). Thirdly, for so-called mixed phase clouds (clouds containing both liquid water and ice), increases in temperature would promote phase changes from ice to liquid, implying a greater proportion of cloud water (\citep{Ceppi2017}). In all cases, the effect will be to increase the cloud optical depth (and hence albedo), implying that an increase in CLWC with warming is a potential \emph{negative} feedback. Modern GCMs corroborate this, typically exhibiting a small, global net negative cloud optical feedback (\citep{Ceppi2017}).

One simple way of measuring the potential strength of this feedback due to the first mechanism above was outlined in \citep{Somerville1985} (see also \citep{Betts1987} and \citep{Charlock2012}), which we will apply here. For clouds within the approximate temperature range $-25^{\circ}C$ to $0^{\circ}C$, both observational data and models show that CLWC increases with increased temperatures, roughly in line with what is predicted by the moist adiabatic approximation. By stratifying clouds according to temperature, we can compute the gradient $\lambda$ of CLWC with respect to changes in temperature, by fitting a straight line to the data. The larger is $\lambda$, the more CLWC would be expected to increase per unit Kelvin. Therefore, the magnitude of $\lambda$ gives a first-order estimate of how strong (i.e. how negative) the cloud optical feedback in the model is.

We estimated this metric for clouds in the FastSPHINX experiments for pressure levels 850, 700hPa, and 500hPa, which span the approximate range for which we find clouds in the right temperature range. The mean value of $\lambda$ across the full deterministic ensemble at 700hPa is $3.62\cdot 10^{-6} kg\cdot m^{-2}K^{-1}$, with a standard error of $0.117\cdot 10^{-6} kg\cdot m^{-2}K^{-1}$. The stochastic mean is $3.81\cdot 10^{-6} kg\cdot m^{-2}K^{-1}$, with a standard error of $0.046\cdot 10^{-6} kg\cdot m^{-2}K^{-1}$. The difference is statistically significant, and represents an increase of around 5\% with SPPT. An almost identical, statistically significant increase is found at the 500hPa level. However, at 850hPa, there is no statistically significant change in the gradient across the two ensembles. This may again be due to the tapering of the stochastic perturbations near the surface, which imply a weakening of the scheme's impact at lower pressure levels.

As noted previously, the difference between the stochastic and deterministic ensemble mean CLWC stays roughly constant in time, despite the stochastic simulations warming less, consistent with the above calculations. While changes in CLWC cannot be easily untangled from changes in cloud cover, already seen to be different for the two sets of simulations, this suggests that SPPT has increased the negative cloud optical feedback. 

Note that because CLWC was not available on levels in the SPHINX simulations, the computation of $\lambda$ could not be carried out for these. We cannot therefore rule out that this change in $\lambda$ may be a feature of the spin-up phase only.

\subsection{Non-linearity of Climate Sensitivity}

There is some evidence that there is a dependence of the amount of global warming (due to some fixed forcing) on the initial global mean temperature (e.g. \citep{Meraner2013, Bloch-Johnson2015, Friedrich2016, Caballero2013}). These studies suggest that the system's sensitivity may increase with increased temperatures, for example due to an increase in the water vapor feedback (\citep{Meraner2013}). Because the stochastic simulations start with a slightly cooler mean state, it is possible that the reduced GHG-sensitivity may be due to this effect.

The exact extent of this dependence on the initial state is not well understood, and so the attribution of a 10\% decrease due to a mean state ${\sim} 0.3K$ cooler cannot be done rigorously. We will therefore simply demonstrate, given one example from the literature, that this hypothesis cannot be immediately discounted. Specifically, \citep{Friedrich2016} use reconstructed paleo-climatic data to estimate climate sensitivity starting from a glacial period or an interglacial period. In the cooler glacial period, they find a sensitivity to doubling CO2 of $1.78K$, while in the warmer interglacial period their estimate increases to $4.88K$. Under the assumption that climate sensitivity scales linearly with temperature, this implies an approximately 55\% increase in sensitivity per Kelvin temperature increase. An $0.3K$ difference would, using these estimates, imply a difference in sensitivity of around 16\%. Consequently, the decreased climate sensitivity with SPPT may be due at least partially to the altered mean state.

\section{Discussion and Conclusions}
\label{sec:conclusions}

We have demonstrated that the inclusion of a stochastic scheme in a coupled climate model can significantly change not only the model's mean surface temperature, but also its GHG-sensitivity. Concretely, in the EC-Earth model considered here, the inclusion of the SPPT scheme led to a cooling of the model's mean surface temperature by ${\sim}0.3K$, as well as decreasing the magnitude of global warming by around 10\%. It is possible that the ultimate difference, after allowing the simulations to reach a new equilibrium, may be larger still. We linked the change in the model's historical climate to a rapid increase in cloud liquid water when turning on SPPT, which we speculated was due to the non-linear impact of perturbations on condensation triggers. A similar increase had been noted also by \citep{Strommen2019}, using a different version of EC-Earth, suggesting this may be a robust impact of the scheme. The increased cloud liquid water amplifies the cloud albedo, resulting in a cooling which is never fully compensated for by subsequent climate feedbacks. For the decrease in climate sensitivity, three possible mechanisms were suggested. Firstly, the stochastic scheme appears to slow down the steep decline in low-level cloud cover observed in the model as the surface temperature increases: this implies a reduction in the positive low cloud cover feedback in the stochastic model. Secondly, the negative cloud optical feedback appears to be slightly amplified, again implying reduced sensitivity. Finally, some role may be played by possible non-linearity of climate sensitivity, e.g. due to the non-linearity of water vapor feedback observed in some studies. 

These results reinforce a growing body of evidence showing that the inclusion of a stochastic scheme can, through non-linear processes, shift a model to a notably different climate attractor. In particular, the choice of a stochastic scheme (including not having one at all), can significantly impact both a GCMs reproduction of historical data, as well as its projections of future climate change. However, some important outstanding questions remain. Most importantly, is the impact on the cloud cover feedback an effect of the changed mean state alone, or is there a more dynamic interaction of SPPT with the cloud thinning processes? In other words, would the models' climate sensitivities be more similar if the stochastic model were tuned in a similar way to the deterministic model? If so, it is possible that the same impact could have been achieved through parameter perturbations in the model: this would require further investigation to answer.

In light of our results, it would be of great interest to assess further the impact of SPPT on the key, small-scale processes driving cloud feedbacks. This could potentially inform the development of new stochastic schemes. Given the ability of such schemes to improve other key aspects of models, this may, ultimately, lead to more accurate projections of global warming.

%
%

\clearpage

\acknowledgments

Free accessibility of the EC-Earth SPHINX data to the climate user community is granted through a dedicated THREDDS Web Server hosted by CINECA (https://sphinx.hpc. cineca.it/thredds/sphinx.html), where it is possible to browse and directly download any of the Climate SPHINX data used in the present work. Further details on the data accessibility and on the Climate SPHINX project itself are available on the official website of the project (http://www.to.isac.cnr.it/sphinx/). The SPHINX data was generated with computing resources provided by CINECA and LRZ in the framework of Climate SPHINX and Climate SPHINX reloaded PRACE projects. 

SSM/I and SSMIS data are produced by Remote Sensing Systems. Data are available at www.remss.com/missions/ssmi.

KS and TNP acknowledge support by the PRIMAVERA project, funded by the European Commission under grant agreement 641727 of the Horizon 2020 research program. PAGW acknowledges European Research Council grant 291406 and Natural Environment Research Council grant NE/P002099/1. The SPHINX model source code, used to generate the FastSPHINX simulations, was generously provided to us by P. Davini (CNR). The FastSPHINX experiments were carried out through the ECMWF Special Project `spgbtpsp'. 

\clearpage


%
%
%
%
%
%
%
%




\listofchanges

\end{document}